\documentclass[conference]{IEEEtran}
\IEEEoverridecommandlockouts

\usepackage{amssymb}
\usepackage{graphicx}
\usepackage{qtree}
\usepackage{circuitikz}
\usepackage{tikz}
\usepackage{mathtools}
\usepackage{tkz-berge}
\usepackage{booktabs}
\usepackage{hyperref}
\usepackage{subcaption}
\usepackage{dirtree}
\usepackage{booktabs}
\usepackage{multirow}

\usetikzlibrary{shapes,snakes}

\usepackage{amsthm}

\theoremstyle{definition}

\usepackage[linesnumbered,ruled,vlined]{algorithm2e}
\usepackage{algpseudocode}

\usetikzlibrary{shapes.gates.logic.US,trees,positioning,arrows}

\usepackage{lipsum}

\usepackage[latin1]{inputenc}
\usetikzlibrary{shapes,arrows}

\usepackage{pgfplots}
\pgfplotsset{width=8.3cm,compat=1.9}

\usepackage{pgfplotstable}
\usepackage{pgfplots}
\usetikzlibrary{patterns}
\pgfplotsset{compat=1.3}
\usepackage{xcolor}
\usepackage{csvsimple}

\usepackage[linesnumbered,ruled,vlined]{algorithm2e}
\usepackage{algpseudocode}

\begin{document}

\title{An Automated Security Analysis Framework and Implementation for Cloud}

\author{\IEEEauthorblockN{Hootan Alavizadeh\textsuperscript{1}, Hooman Alavizadeh\textsuperscript{2}, Dong Seong Kim\textsuperscript{3}, Julian Jang-Jaccard\textsuperscript{2} and\\ Masood Niazi Torshiz\textsuperscript{1}}
		\IEEEauthorblockA{\textsuperscript{1} Computer Engineering Department,\\ Imam Reza International University, Mashhah, Iran.\\
		\textsuperscript{2} School of Natural and Computational Sciences,\\ Massey University, Auckland, New Zealand.\\
		\textsuperscript{3} School of Information Technology and Electrical Engineering,\\ The University of Queensland, Australia.\\
		Email: h.alavizadeh@imamreza.ac.ir, niazi@mshdiau.ac.ir\\ \{h.alavizadeh, j.jang-jaccard\}@massey.ac.nz and dan.kim@uq.edu.au}
}

 \maketitle

	\begin{abstract}
		Cloud service providers offer their customers with on-demand and cost-effective services, scalable computing, and network infrastructures. Enterprises migrate their services to the cloud to utilize the benefit of cloud computing such as eliminating the capital expense of their computing need. There are security vulnerabilities and threats in the cloud. Many researches have been proposed to analyze the cloud security using Graphical Security Models (GSMs) and security metrics. In addition, it has been widely researched in finding appropriate defensive strategies for the security of the cloud. Moving Target Defense (MTD) techniques can utilize the cloud elasticity features to change the attack surface and confuse attackers. Most of the previous work incorporating MTDs into the GSMs are theoretical and the performance was evaluated based on the simulation. In this paper, we realized the previous framework and designed, implemented and tested a cloud security assessment tool in a real cloud platform named UniteCloud. Our security solution can (1) monitor cloud computing in real-time, (2) automate the security modeling and analysis and visualize the GSMs using a Graphical User Interface via a web application, and (3) deploy three MTD techniques including Diversity, Redundancy, and Shuffle on the real cloud infrastructure. We analyzed the automation process using the APIs and showed the practicality and feasibility of automation of deploying all the three MTD techniques on the UniteCloud.
	\end{abstract}

\section{Introduction}
	The growth of the cloud computing as a powerful and affordable context for users has caused many business and commerce migrate to this on-demand, scalable, and cost-effective paradigm. The organizations outsource their network infrastructures, computing needs, software and services into the cloud in order to benefit from the cloud's utilities such as economical benefits (cutting off physical resources and damages). However, many organizations and enterprises find this migration undesirable due to security issues in the cloud~\cite{sgandurra2016evolution, zissis2012addressing}.
	
	Many security mechanisms and defensive strategies have been proposed by researchers both theoretically and practically. In order to improve the security of cloud computing, it is important to evaluate the security posture of cloud. Graphical Security Models (GSMs) (such as Attack Graphs (AGs)\cite{cook2016scalable}, Attack Trees (ATs)~\cite{kordy2013dag}, Attack-defense threes (ADTrees)\cite{kordy2014attack}, HARMs~\cite{hong2012harms}) are the widely adopted methods to analyze the security of enterprise networks~\cite{granadillo2012individual,jia2015towards}; a GSM can be used to define attack surfaces and summarize the attack scenarios, and compute security metrics. 
	    Morever, GSMs can be used to evaluate the cloud security posture. GSMs can also be used to evaluate the effectiveness of defensive techniques such as Moving Target Defense (MTD). MTD techniques are proactive defensive techniques and the primary idea is mainly changing the attack surface in order to introduce confusions to attackers carrying out cyber attackers. There are a few researches in this line. Only a few researches have been proposed for the uses of GSM in evaluating MTD techniques for cloud computing. However, most of the previous researches are theoretical and use simulation only~\cite{hong2016assessing, alavizadeh2018evaluation, alavizadeh2017effective, alavizadeh2018comprehensive} to show the feasibility of their approaches.\\
	    To the best of our knowledge, the incorporation of GSMs and MTD techniques together for security analysis and deployment of MTD techniques in the infrastructures of the real clouds has not been proposed.

	In this paper, we tackle the aforementioned shortcomings by designing and development of a cloud security assessment framework which can automatically monitor, model, and analyze a private cloud security and deploy the MTD techniques on the cloud infrastructures. In this paper, we focus on the practical side rather theoretical appraisal. We demonstrate the practicality of implementation, feasibility of automation, usability of the project using a real cloud platform named UniteCloud~\cite{he2016reverse, pang2017cdmc}.
	
	The main contributions of this paper are summarized as follows:
	
	\begin{itemize}
		\item{\textit{Cloud monitoring}: We developed a cloud security framework which can automate the process of cloud vulnerability scanning in order to collect the information of the cloud's components together with the vulnerabilities of each component.} 
		\item{\textit{Cloud security evaluation:} Cloud security framework can create the HARM based on the collected information for security analysis and MTD evaluation purposes.}
		\item{\textit{MTD Deployment}: Cloud security framework automated the deployment of three MTD techniques such as Diversity, Redundancy, and Shuffle on the real cloud infrastructure.}
		\item {\textit{Automation evaluation}: We investigated on a private cloud platform and uses of OpenStack Application Programming Interfaces (APIs) to analyze the automation process for implementation steps.}
		\item {\textit{Automation evaluation}: We developed a graphical user interface (GUI) as a web application for interaction between cloud security framework and security experts including both cloud provider view and HARM~\cite{hong2016assessing} visualization.}
	\end{itemize}
	
	

	The rest of the paper is organized as follow. Section \ref{sec:pro} defines the proposed approach including a brief explanation on preliminaries, concepts, and definitions. Section \ref{sec:design} presents the design and implementation of the cloud security assessment framework. Discussion and limitations of this work are given in Section \ref{sec:discussion}. Section \ref{sec:RW} summarizes the related work. Finally, we conclude the paper in Section \ref{sec:conclusion}.
	
	
\section{Proposed Approach}\label{sec:pro}
In this work, we implement a cloud security assessment framework which is able to monitor the cloud, analyze, and deploy the three MTD techniques including Shuffle, Diversity, and Redundancy on the real infrastructures of the cloud. The main part of this paper is the automation of the cloud assessment framework in the real cloud. The uses of APIs in the implementation and automation of the project are nontrivial. Automation process needs a deep understanding of the infrastructures and platforms in which the private cloud uses. On the other hand, the security analysis framework should be able to handle the cloud constraints defined by both cloud provider and security experts such as uses of services, access to controllers, \text{etc.}. This work includes four main phases elucidated as follows. (1) Information Collection, (2) Cloud Security Modeling using HARM, (3) Security Analysis Engine, (4) Deployment Phase.

\subsubsection{Preliminaries}\label{sec:pre}
In this section, we describe the related concepts and definitions used throughout this paper. We first define a running example as the main scenario for the migration of enterprises to the cloud. Later on, we deploy the proposed scenario together with the proposed approach on the UniteCloud. 
\subsection{Running Example}
\begin{figure}[t]
	\centering
	\includegraphics[height=5.5cm,width=9cm]{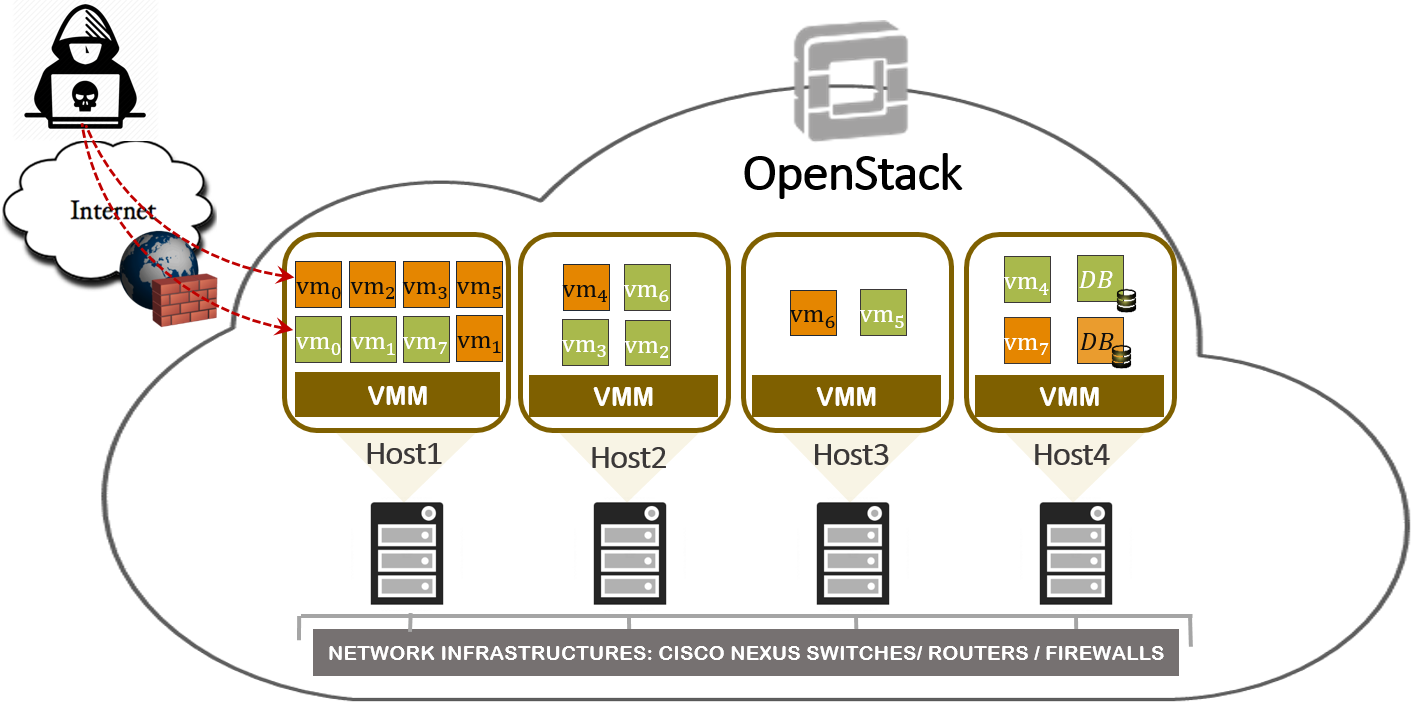}
	\caption{Running Example: a Cloud example including the different Hosts and Virtual Machines (VMs) of two organizations.}
	\label{fig:running example}
\end{figure}

\begin{figure*}[t]
	\centering
	\includegraphics[height=9.5cm]{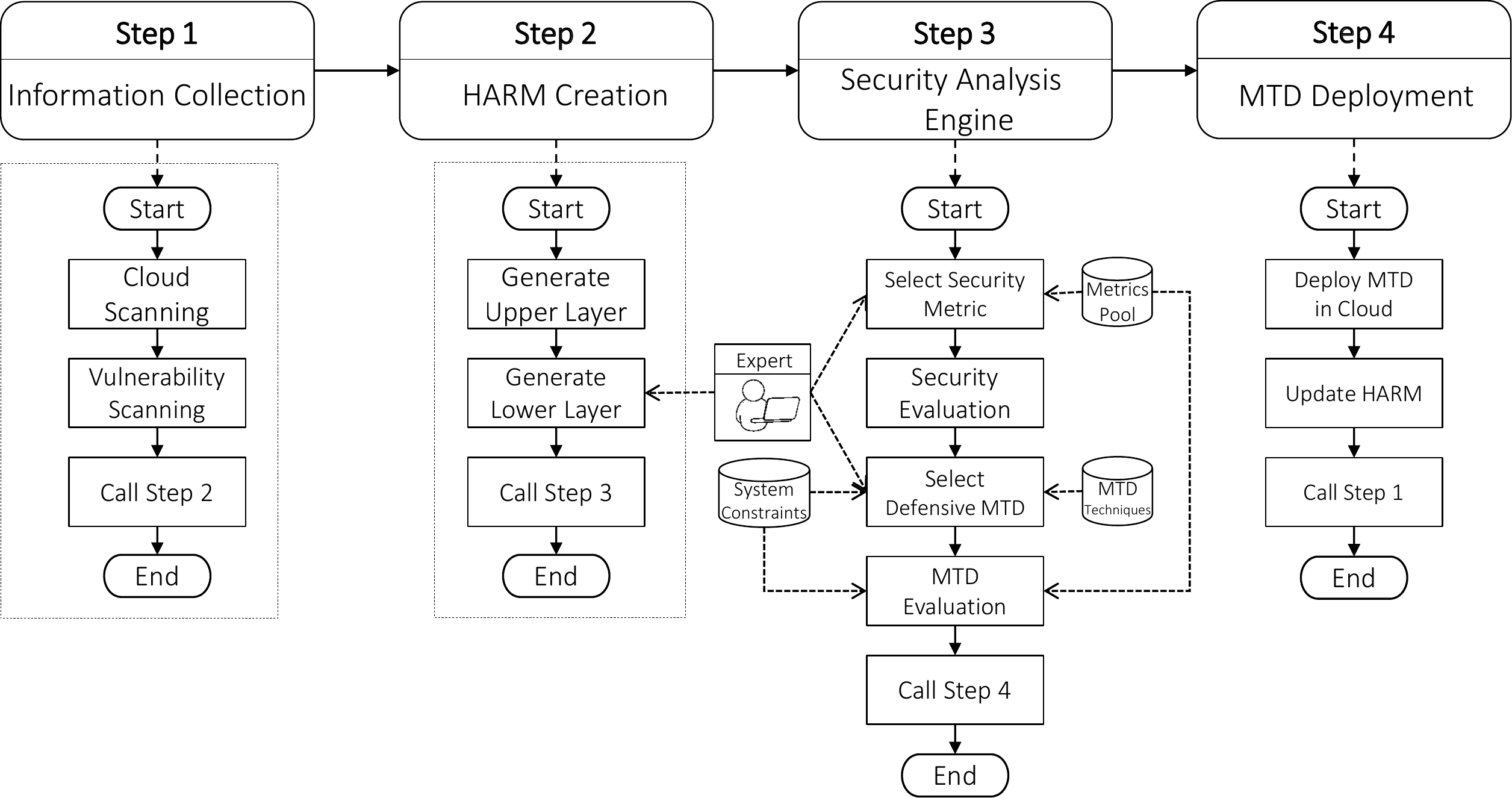}
	\caption{Security Modeling, Analysis, and Deployment Phases}
	\label{fig:model}
\end{figure*}

Fig.~\ref{fig:running example} shows the running example scenario on the migrations of two independent organizations entitled Enterprise-1 ($\text{EP}_1$) and Enterprise-2 ($\text{EP}_2$) to a private cloud. Those companies decide to cut off the physical equipment and use a private cloud for accommodating their computing needs. Each organization has launched 8 Virtual Machines (VMs) on the cloud together with a Database (DB) creating a virtual network. We assume that the first four VMs use Windows10 instances and the rest does Linux iUbuntu. Moreover, the $vm_0$ for both organizations are connected to the Internet. Later on, we deploy the running example shown in Fig.~\ref{fig:running example} in the UniteCloud.

\subsubsection{System model}
System constraints are usually defined based on both cloud provider and security experts. For instance, the cloud provider can determine which cloud zones or physical hosts are available for the customers. Moreover, the cloud provider can set the limitations on the physical hosts such as defining the maximum VMs can be located on each host and \textit{etc.}. The system constraints are defined due to different reasons like workload balance or energy saving, security purposes~\cite{han2017using}. On the other hand, the security experts of enterprises migrated into the cloud may have their own security policies like defining firewalls rules and Access Control Lists (ACL).

\subsubsection{Attack Model}
We assume that an attacker can launch the attacks from outside of the cloud using exploiting the software vulnerabilities of the VMs connected to the Internet. Then, the attacker can launch a series of other attacks in order to access the DB along the identified attack paths. 

\subsection{Security Modeling}
In this paper, we use HARM~\cite{hong2012harms,hong2016assessing} for graphical security modeling and analysis and MTD evaluation. HARM consists of two hierarchical layers which use an AG in the upper layer and an AT in the lower layer. The upper layer captures the reachability of nodes (in here, VMs) and the lower layer holds the vulnerabilities information. Security modeling is the first phase of the cloud security framework. Security modeling consists of two phases: (1) information collection, (2) HARM creation shown as steps 1 and 2 in Fig.~\ref{fig:model}. First, the cloud infrastructure should be scanned in order to obtain Hosts, VMs, and reachability information. Then, the vulnerabilities existing on each VM should be obtained using the vulnerability scanning tools~\cite{nessus}. Information gathering is a crucial phase for the security modeling. In the next step, an HARM can be constructed using the obtained information. The reachability information can be used to generate the upper layer of HARM where an AG is used to show all the possible attack scenarios given system and attack model. Moreover, the vulnerabilities information can be used to construct the lower layer of HARM which uses the ATs. However, generating ATs from vulnerabilities needs a clear understanding of the vulnerabilities and the way in which they can be exploited. For instance, an attacker can exploit only one vulnerability to penetrate into a VM, or the attacker may need to exploit a set of vulnerabilities to penetrate into a VM. In the former case a logical OR-gate can be used and for the latter, a combination of logical AND/OR-gates can be used~\cite{roy2012attack}. The uses of logical AND/OR-Gates and computation approach are presented in~\cite{roy2012attack}. However, a security expert can help to define the vulnerabilities relations. Entry points of the cloud are actually the VMs connected to the Internet. Those VMs are the entry points of the attacks as well. The target could be any VM which includes important information or runs crucial services. We assume that the DB is the attackers' target. Both entry points and target are captured in the upper layer of the HARM. The upper layer of HARM can be generated using reachability information obtained from analyzing the firewall rules.

\subsection{Security Analysis}\label{subsec:SA}
Security analysis engine has two main phases: general security evaluation and MTD evaluation. In order to assess the cloud security, the security experts can choose or prioritize the use of the security metrics based on their security requirements. One of the most important metrics is cloud risk value which shows the overall (system) risk associated with the cloud. Other security metrics like Attack Costs, the probability of attack success, Mean Time to Attack (MTTA),\text{etc.}~\cite{yusuf2016security} can also be added to a set (or pool) of the security. HARM can be adopted based on the security metrics pool and compute the security metrics for analysis. It uses the values of each vulnerability represented at the lower layer of HARM such as Impact, Exploitability, CVE CVSS Base Score and uses them through a bottom-up approach.
The detailed explanation of the security metrics and the calculation steps is given in~\cite{yusuf2016security}.

\section{Design and Implementation}\label{sec:design}
This section provides the design and development of a security analysis framework for cloud computing. We investigate the feasibility and practical requirements such as Software tools, packages, programming interfaces, libraries in order to implement and automate the security analysis tool and MTD techniques in the real-world cloud deployment. We develop a framework which can perform security modeling, evaluation, MTD deployment for enterprises migrated into the private clouds. The cloud security framework is able to automate information collection: cloud scanning, vulnerability scanning, HARM creation, security evaluation, and MTD deployment on a real cloud infrastructure. To implement the framework we utilize a private cloud named UniteCloud and develop our framework on UniteCloud as a case study. However, we believe that our developed framework can be adopted for other private clouds as well.
\subsection{Case Study: UniteCloud Analysis}

The UniteCloud uses the OpenStack cloud platform. For setting up the project, we can either use OpenStack horizon dashboard or utilize OpenStack APIs. The setup process includes the creation of VMs with different flavors and OS, assigning internal and floating IP addresses, defining firewall rules and ACL, \textit{etc}. However, we first create the cloud example VMs shown in Fig.~\ref{fig:running example} into the UniteCloud infrastructures using the horizon dashboard. Further, we utilize the OpenStack APIs for automation. Table~\ref{setup} shows the created VMs including the related information on each host. The cloud consisting of 16 physical Hosts (Compute Hosts) is distributed over three availability zones: IBMZone, HPZone, and Nova. However, in this paper, we used four hosts each of which includes different VMs. We assign two flavors for the VMs: m1.medium and m1.generic. The specification of the former VM is 2 VCPUs, 4 GB RAM, and 80 GB Disk, and that of the latter is 1 VCPUs, 1 GB RAM, and 20 GB Disk, respectively. We assign two floating IPs for both $vm_0$ and $vm_1$ of two enterprises denoting the entry points of the cloud. Moreover, both $vm_7$ of enterprises are connected to the DB.
\begin{table}[t]
	\caption{Configuration and setup for VMs and hosts in the cloud. Note: floating IPs are denoted as asterisked}
	\label{setup}
	\begin{tabular}{@{}lllll@{}}
		\toprule
		\begin{tabular}[c]{@{}l@{}}Host \& \\Zone Name\end{tabular}              & \begin{tabular}[c]{@{}l@{}}VM \\ Name\end{tabular} & OS          & \begin{tabular}[c]{@{}l@{}}IP \\ Addresses\end{tabular}               & \begin{tabular}[c]{@{}l@{}}Flavor\\ Size\end{tabular} \\ \midrule
		\multirow{8}{*}{\begin{tabular}[c]{@{}l@{}}$h_0$\\ IBMZone\end{tabular}} & $vm_1$-EP2                                         & Win10       & \begin{tabular}[c]{@{}l@{}}$172.16.7.33$\\ $192.168.1.100^*$\end{tabular}                                                           & m1.medium                                             \\
		& $vm_2$-EP2                                         & Win10       & $172.16.7.32$                                                           & m1.medium                                             \\
		& $vm_5$-EP2                                         & Ubuntu14.04 & $172.16.7.39$                                                           & m1.generic                                            \\
		& $vm_3$-EP2                                         & Win10       & $172.16.7.36$                                                           & m1.medium                                             \\
		& $vm_7$-EP1                                         & Ubuntu14.04 & $172.16.19.16$                                                          & m1.generic                                            \\
		& $vm_0$-EP1                                         & Win10       & \begin{tabular}[c]{@{}l@{}}172.16.19.14\\ $192.168.1.239^*$\end{tabular} & m1.medium                                             \\
		& $vm_0$-EP2                                         & Win10       & \begin{tabular}[c]{@{}l@{}}172.16.7.35\\ $192.168.1.149^*$\end{tabular}  & m1.medium                                             \\
		& $vm_1$-EP1                                         & Win10       & \begin{tabular}[c]{@{}l@{}}$172.16.19.12$\\ $192.168.1.63^*$\end{tabular}                                                          & m1.medium                                             \\ \midrule
		\multirow{4}{*}{\begin{tabular}[c]{@{}l@{}}$h_1$\\ IBMZone\end{tabular}} & $vm_4$-EP2                                         & Ubuntu14.04 & $172.16.7.37$                                                           & m1.generic                                            \\
		& $vm_6$-EP1                                         & Ubuntu14.04 & $172.16.19.18$                                                          & m1.generic                                            \\
		& $vm_2$-EP1                                         & Win10       & $172.16.19.15$                                                          & m1.medium                                             \\
		& $vm_3$-EP1                                         & Win10       & $172.16.19.11$                                                          & m1.medium                                             \\ \midrule
		\multirow{2}{*}{\begin{tabular}[c]{@{}l@{}}$h_2$\\ HPZone\end{tabular}}  & $vm_6$-EP2                                         & Ubuntu14.04 & $172.16.7.40$                                                           & m1.generic                                            \\
		& $vm_5$-EP1                                         & Ubuntu14.04 & $172.16.19.19$                                                          & m1.generic                                            \\ \midrule
		\multirow{2}{*}{\begin{tabular}[c]{@{}l@{}}$h_3$\\ HPZone\end{tabular}}  & $vm_7$-EP2                                         & Ubuntu14.04 & $172.16.7.38$                                                           & m1.generic                                            \\
		& $vm_4$-EP1                                         & Ubuntu14.04 & $172.16.19.17$                                                          & m1.generic                                            \\ \bottomrule
	\end{tabular}
\end{table}
\begin{figure}[b]
	\centering
	\includegraphics[height=4.75cm, width=9cm]{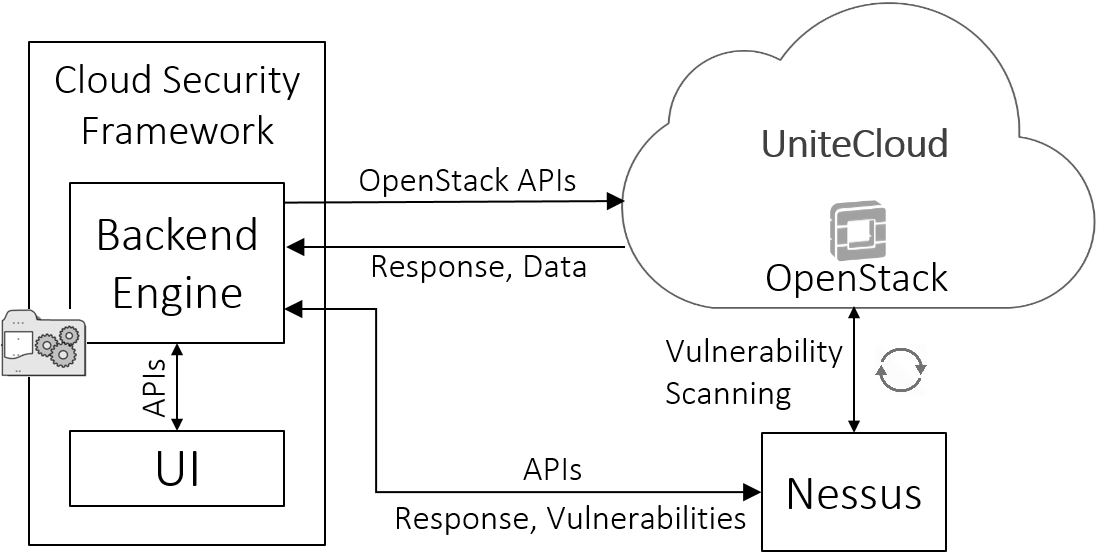}
	\caption{Security framework and communication overview.}
	\label{fig:Framework}
\end{figure}
\begin{figure}[b]
	\centering
	\includegraphics[height=4.75cm, width=9cm]{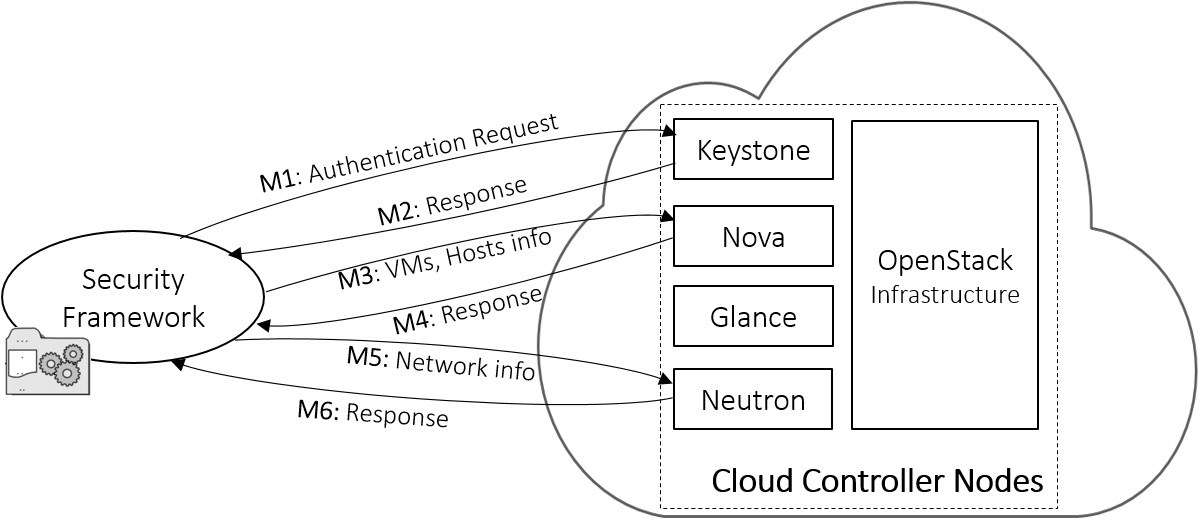}
	\caption{OpenStack API Calls for information collection phase.}
	\label{fig:API-calls}
\end{figure}
\subsection{Cloud Security Framework}

The cloud security framework utilizes the following programming languages, tools, and concepts: .NET Core, JSON, JavaScript, jQuery Ajax, Python, Nessus \cite{nessus}, Data-Driven Documents JavaScript (D3.js). In this section, we show the implementation of the cloud security framework. An overview of the prototype and related tools and communication is shown in Fig.~\ref{fig:Framework}. The security framework consists of a backend engine and user interface (UI). The backend engine is responsible for information collection, security modeling, analysis, and deployment phases which are demonstrated in Fig.~\ref{fig:model}. The UI is used for interactions between security experts of enterprises and the backend engine for configuration and visualization purposes. The generated graphical security model can be visualized in the UI.

\subsubsection{Information Collection Automation}
As stated earlier, cloud infrastructure information including VMs and hosts, and the reachability of VMs are required for constructing the upper layer of HARM, and vulnerabilities associated to each VM are required to create the lower layer of HARM. Cloud security framework needs to automatically fetch two information: (1) cloud information such as the number of VMs, the number of physical hosts, the host of each VM, the reachability between the VMs and (2) vulnerabilities information existing on each VM. We use .NET Core as the backend engine programming language and call APIs in order to access both OpenStack and Nessus automatically and fetch information. Accessing to the UniteCloud OpenStack consists of two parts: OpenStack authentication and fetching information. OpenStack uses Keystone feature for user authentication. Moreover, it uses nove-computes, neutron-networks, Glance-images features for different purposes such as accessing to compute nodes (VMs, Hosts, Zones, \textit{etc}). In order to access to the OpenStack and retrieve the information, we first need to access keystone using APIs for authentication. The username, password, and domain name are sent to the Keystone controller using a JSON API call for authentication. Once the user is authenticated using the Keystone authenticaion method, OpenStack sends a response including the authentication token (X-Subject-Token), other OpenStack Controllers' address including nova, neutron, glance, cinder, \textit{etc.} which can be used for further API calls. The received message should be first parsed to receive the authentication token together with the nova controller address. Then, the backend engine sends another API call using the authentication token and the nova controller to gather the list of VMs and Hosts. The received message contains unnecessary/irrelevant information including VM status, availability zone, created and updated time, \textit{etc.}, the message should be parsed to fetch only the required information. Similarly, another API including the authentication token and neutron controller should be called to get network-related information. The received information should be again parsed to obtain VMs' IP addresses and the reachability of VMs.
Fig.~\ref{fig:API-calls} demonstrates the API calls and related responses between the cloud security framework and OpenStack in order to gain the information. Beside the VMs and reachability information, we need vulnerabilities information for each VM on the cloud. We use Nessus~\cite{nessus} to scan the cloud and obtain vulnerabilities. In the next step, cloud security framework uses a backend engine to access to Nessus and retrieve the vulnerabilities' information. the first API called is used for authentication. Having obtained the response message, the backend engine sends other API calls using the authentication token in order to get the vulnerability information. the extracted information contains useful information related to Vulnerability, possible threats, Base Score\cite{mell2006common}, severity, \textit{etc.}, CVE identifier (CVE-ID). However, cloud security framework only need CVE-ID for selected vulnerabilities so that it can obtain the other information such as vulnerability impact and exploitability through National Vulnerability Database (NVD)\cite{mell2006common}. The pseudocode for the overall information collection is shown in Algorithm~\ref{alg:IR}. Note that the measured time is only for API calls and responses times, and the cloud vulnerability scanning time is not measured. However, cloud scanning using Nessus is a time-consuming process and cannot be done frequently. Instead, it can be run once a while to keep the vulnerabilities updated, or run once a change catches on the VMs such as adding new VM, or changing OS, services, \textit{etc.}

\begin{algorithm}[t]
	\SetAlgoLined
	\DontPrintSemicolon
	{\footnotesize \tcc*[f]{Input info. needed for OpenStack (Ops)}}\\
	\KwData{Ops-user-credential, Keystone-Controller-Url}
	{\footnotesize \tcc*[f]{Input Info. needed for Nessus (NS)}}\\
	\KwData{NS-user-credential, Nessus-Session-Url}
	{\footnotesize \tcc*[f]{Result: Dictionaries of VMs and Reachability, VMs and Vulnerabilities}}\\
	\KwResult{VMs\_Links\_Dic, VMs\_Vuls\_Dic}
	\Begin{\small
		{\footnotesize \tcc*[f]{Cloud Scanning: fetch Host \& VM info.}}\\
		Credential-Data$\leftarrow$JSonConvert(Ops-user-credential)\;
		{JResult$\leftarrow$\footnotesize API\_Call(Credential-Data, Keystone-Controller-Url)}\;
		Auth\_Token$\leftarrow$Parse(JResult, Authentication)\;
		{\footnotesize Controllers\_List$\leftarrow$Parse(JResult, Nova and Neutron Controllers)}\;
		{\footnotesize Host-VM-Info$\leftarrow$ API\_Call(Auth\_Token, Nova-Controller)}\;
		{\footnotesize Network-Info$\leftarrow$ API\_Call(Auth\_Token, Neutron-Controller)}\;
		{\footnotesize \tcc*[f]{Parsing and saving the fetched data}}\\
		{\textit{Hosts\_List}$\leftarrow$}Parse(Host-VM-Info, Hosts)\;
		{\textit{VMs\_List}$\leftarrow$}Parse(Host-VM-Info, VMs)\;
		{\textit{Reachability\_List}$\leftarrow$}Parse(Network-Info, Reachability)\;
		{\footnotesize VMs\_Links\_Dic $=$ Create Dictionary[VM, VM]}\;
		{\footnotesize \tcc*[f]{Nessus Scanning: fetch vulnerabilities}}\\
		Credential-Data$\leftarrow$JSonConvert(NS-user-credential)\;
		{JResult$\leftarrow$\footnotesize API\_Call(Credential-Data, Nessus-Session-Url)}\;
		Auth\_Token$\leftarrow$Parse(JResult, token)\;
		{\footnotesize JResult$\leftarrow$\footnotesize API\_Call(Auth\_Token, Nessus-vulnerabilities-Url)}\;
		{\footnotesize VMs\_Vuls\_Dic $=$ Create Dictionary[VM, Vulnerabilities-List]}\;
		{\footnotesize \tcc*[f]{Get \& save vulnerabilities on each VM}}\\
		\ForEach {$vm \in \text{VMs\_List}$}{
			\textit{Vuls-Info}$\leftarrow$Parse(JResult, vm)\;
			Add $vm$ and \textit{Vuls-Info} into VMs\_Vuls\_Dic\;
		}
		{\footnotesize \tcc*[f]{Return reachability of VMs~~~~~~~~~~~}}\\
		\KwOut{VMs\_Links\_Dic}
		{\footnotesize \tcc*[f]{Return vulnerabilities on each VM~~~~}}\\
		\KwOut{VMs\_Vuls\_Dic} 
	}
	\caption{Information Collection Procedure}
	\label{alg:IR}
\end{algorithm}
\subsubsection{HARM Creation}
The upper layer of HARM can be generated using the VMs and reachability information obtained from the previous step. This information is saved as a key and value dictionary representing the VMs' links as a graph. Thus, the backend engine can generate the AG based on the dictionary. The second part of the information obtained from Nessus scanning is a dictionary of VMs and related vulnerabilities on each VM which can be used to generate the lower layer of HARM. The lower layer of HARM uses the ATs. The backend Engine uses Python programming language to generate HARM. However, other software and tools can also be used like Gephi which is a network analysis and visualization software package. The reason why we use Python is the NetworkX feature which is a useful package for network graph generation. Moreover, we use Python~\cite{jia2015towards} as the security analysis engine to compute security metrics and evaluate MTD techniques. 

\subsubsection{Security Analysis Engine}
Security analysis engine is implemented on the backend engine using Python. It consists of security evaluation and MTD evaluation subroutine. Security analysis engine uses the generated HARM and the security metrics. Actually, security experts can choose various security metrics and add them to the metric pools such as System Risk, Attack Cost, MTTA, Attack Success Probability. Once the security metrics are selected, security analysis engine uses HARM for security evaluation and computing the selected security metrics. Security analysis framework uses MTD techniques as the main defensive strategies for security the organisations on the cloud. However, deploying MTD techniques could be limited based on system constraints. For instance, VM-LM (Shuffle technique) might be restricted from one host to another one due to lack of space on the target host, or OS Diversification (a Diversity technique) could be limited to only a few OS instances due to the cost of the licence for the cloud provider. Thus, the MTD techniques should be chosen based on the defined system constraints.

\subsubsection{MTD Deployment Implementation}
\begin{figure}[t]
	\centering
	\includegraphics[height=3.85cm]{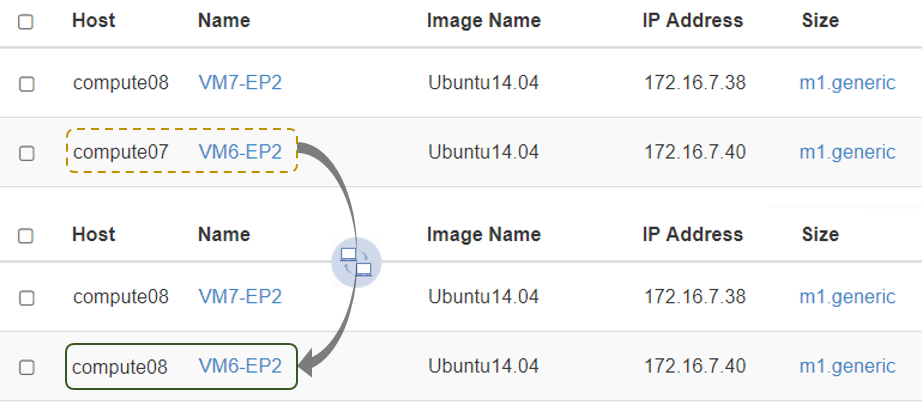}
	\caption{VM-LM: Migration of$vm_6$-EP2 from Compute07 to Compute08.}
	\label{fig:Shuffle}
\end{figure}

The final phase of the cloud security framework is the deployment of selected MTD techniques on the cloud infrastructure. It uses .NET Core and OpenStack APIs to deploy MTD techniques, it utilizes glance for creating and retrieving OS instance images, nove, and network controllers for accessing and manipulating VMs and Network purposes.
\begin{figure}[t]
	\centering
	\includegraphics[height=3.3cm]{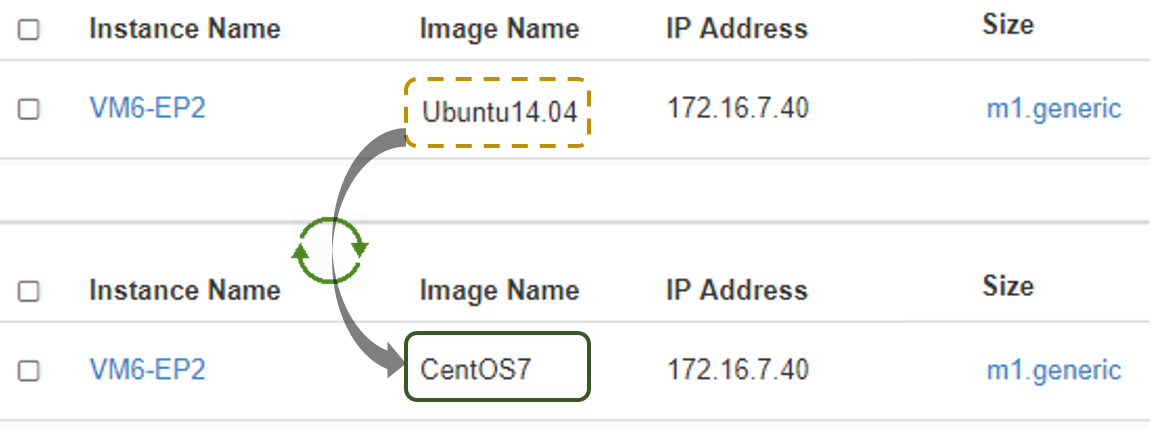}
	\caption{OS Diversification: Ubuntu14.04 replaces with CentOS7 for $vm_6$-EP2.}
	\label{fig:Diversity}
\end{figure}
\paragraph{Diversity}
Security analysis framework uses OS diversification technique for deploying Diversity. In order to deploy Diversity technique, backend engine uses nove to access the desired VM and update the VM instances with another OS image. Similar to the information collection phase, the user credential information should be sent to the Keystone controller using JSON API call for authentication. Backend engine omits this phase as the authentication token is already received in information collection phase; moreover, both nove and glance controllers are fetched from the response message. Before calling API to change the VM instance, we need to fetch the ImageRef by sending an API to glance. Once the response received, the ImageRef associated to the desired VM image can be obtained. Finally, an API should be called to pass the authentication token, VM ID, ImageRef to the nova in order to rebuild the VM with another OS variant. Fig~\ref{fig:Diversity} shows the results of calling APIs for replacing Ubuntu14.04 with CentOS7 for $vm_6$-EP2 on the cloud. Note that Diversity preserves the VM's physical host 
\paragraph{Redundancy}
\begin{figure}[b]
	\centering
	\includegraphics[height=3.45cm]{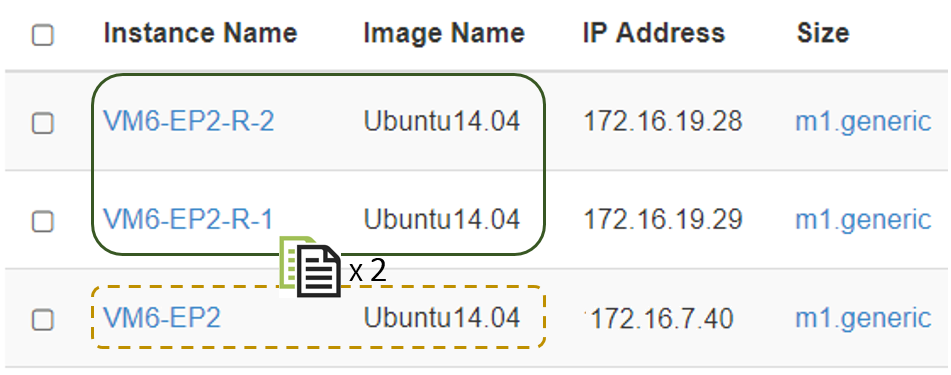}
	\caption{OS Replication: Create 2 replicas for $vm_6$-EP2.}
	\label{fig:Redundancy}
\end{figure}
Based on the Redundancy definition, different replicas of a VM should be created so that each replica has the same feature as the main VM. For instance, the replicated VMs should have the same OS, Flavor, inbound and outbound links from/to other VMs, and should be located on the same physical host. The only difference is the newly assigned IP addresses. Backend engine is responsible for deploying redundancy. However, the number of replicas for deploying redundancy is chosen by either MTD evaluation part or expert entry using UI. There is no feature on OpenStack to create replication for each VM. Thus, deploying redundancy on OpenStack needs creation $r$ new VMs based on the similar existing instance or copied snapshot. Backend engine can use the same authentication token already obtained from the information collection phase and use nova controller. Thus, the backend engine sends an API to nove controller including the authentication token, ImageRef, FlavorRef, NetworkID together with a max\_count which is the number of required replicas ($r$). Fig~\ref{fig:Redundancy} demonstrates the results of calling APIs for the creation of two new replicas of $vm_6$-EP2 with the same OS, links, hosts, flavors, but different IP addresses.
\paragraph{Shuffle}
In this framework, VM-LM is used as the Shuffle technique. VM-LM can be deployed on the OpenStack using nova controller. Similar to other MTD techniques, the backend engine omits the authentication API call because the authentication token and nove controllers have already been fetched in the information collection section. The target host can be selected either by MTD evaluation results or security experts. In order to deploy VM-LM, an API including authentication token together with the VM ID and Target Host ID is called. Fig~\ref{fig:Shuffle} demonstrates the results of calling APIs for migration of $vm_6$-EP2 from Compute07 to Compute08.

\subsection{User Interface (UI) Implementation}
\begin{figure}[h]
	\centering
	\includegraphics[height=7.3cm]{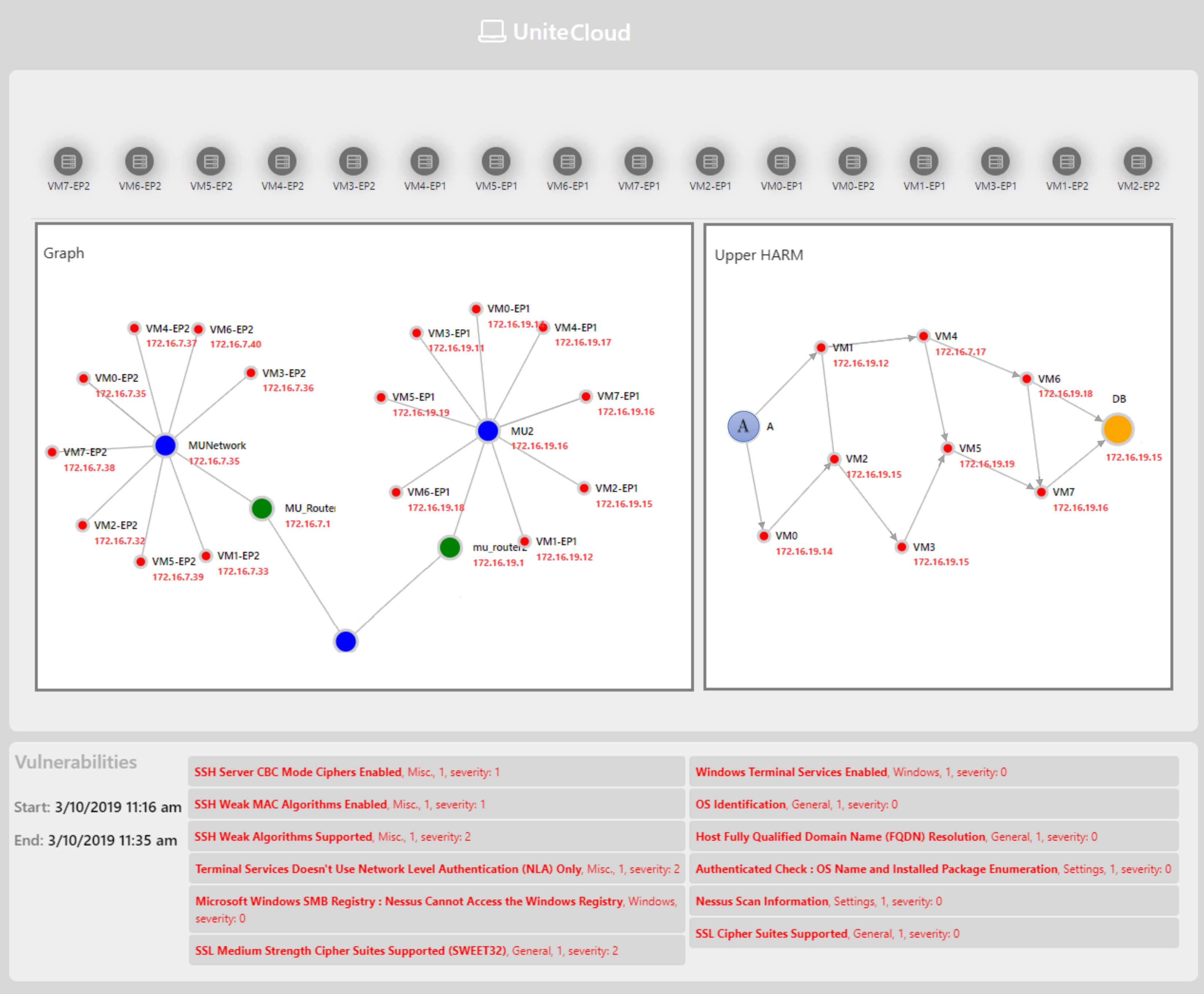}
	\caption{Cloud security framework UI panel: UniteCloud Graph view and HARM visualization.}
	\label{fig:tool}
\end{figure}
Cloud security framework uses a UI in order to interact between the security experts of enterprises and backend engine. Security experts can add update the security metrics pool, choose MTD techniques, analyze and monitor the cloud security using visualization panel. UI is implemented as a web application using JavaScript, JSON, jQuery Ajax, and D3.js interacting with the backend engine. UI web application includes two different perspectives for visualization. Cloud provider and security model previews. Cloud provider preview illustrates the internal connection of the VMs, routers, subnets, and \text{etc.} in the cloud, and security model preview visualizes the generated upper layer of HARM which captures the reachability of VMs based on the firewall rules and possible attack scenarios. UI also shows the vulnerabilities captured for each VM . UI uses internal APIs to communicate with backend engine and update and gain information. Fig.~\ref{fig:tool} demonstrates the UI panel showing two different previous based on the UniteCloud network and HARM view.

\section{Discussion and Limitation}\label{sec:discussion}
Backend engine is the base of the cloud security framework which use the OpenStack APIs to create security, perform security analysis, and deploy MTD techniques. The backend engine is responsible for automating the information collection APIs and MTD deployment APIs. The feasibility and practicability of implementing the backend using OpenStack API calls is important. We evaluated the usability of the backend engine by considering the API calls passing through the backend and two other parties: Nessus vulnerability scanning tool, and OpenStack controllers. The details of API calls like the type of APIs and elapsed times are elucidated in this section.
\paragraph{APIs Automation Evaluation}
Table~\ref{info APIs} shows the details of API calls including the required fields, received messages together with the elapsed time for each call and response. Cloud security framework uses two types of APIs which can be categorized as \textit{informative calls} and \textit{operational calls}. The first group can be only used to get the information like getting authentication tokens, list of hosts, VMs, \textit{etc.}, these APIs will not make any changes on the cloud. Unlike the first group, \textit{operational calls} can perform an operation on the cloud and make the changes such as migrating a VM from one host to another one, or changing the VM's instance, \textit{etc}. Obviously, the response time (RT) for \textit{informative calls} are deterministic. It means the response time of a keystone authentication call is the time elapsed between calling API and receiving the response from keystone. However, \textit{informative calls} consist of: (1) an RT which is the time between calling API and receiving the response (the response can be an acknowledge for the requested API such as accepted, denied, abort, \textit{etc.}), (2) operational time (OT) which means the difference between the start of an operation using API calls and the time in which the task is fully done. For instance, the total time between the start of a VM-LM process and the end of the process. Usually OT is greater than RT. The total RT for \textit{informative calls} passed through backend engine is around $2562$ milliseconds (\textit{ms}). Nessus scanning APIs is also categorized as \textit{informative calls} for authentication and vulnerability information fetches. The total measured RT for Nessus scanning APIs is about $4509$ \textit{ms}. Note that, cloud scanning using Nessus servers is a separate process and is not included in Nessus scanning APIs. Deploying MTD API calls fall into both \textit{informative} and \textit{operational calls}, but as OT is greater than RT the \textit{informative calls} covers \textit{operational calls}. The OT for MTD techniques are asterisked in Table~\ref{info APIs}.
\begin{table}[h]
	\caption{API JSON calls and related information including the (RT). Note: the asterisked times are (OT) }
	\label{info APIs}
	\begin{tabular}{@{}p{0.31cm}p{0.73cm}p{5.73cm}p{0.80cm}@{}}
		\toprule
		\begin{tabular}[c]{@{}l@{}}API\\ Calls \end{tabular} & Type     & Content                                                 & \begin{tabular}[c]{@{}l@{}}Time\\ (ms)\end{tabular} \\ \midrule
		\multicolumn{4}{c}{Cloud Scanning APIs}                                                                                                                                                    \\ \midrule
		M1                                                         & Request  & {[}User, Password, Domain{]}, {[}Keystone Controller{]} & $356$                                                 \\
		M2                                                         & Response & {[}Authentication Token, Controllers' URL{]}            &                                                     \\
		M3                                                         & Request  & {[}Authentication Token{]}, {[}Nova Controller{]}       & \multirow{2}{*}{$1997$}                               \\
		M4                                                         & Response & {[}List of Servers, Hosts, Zones, \textit{etc.}{]}               &                                                     \\
		M5                                                         & Request  & {[}Authentication Token{]}, {[}Neutron Controller{]}    & \multirow{2}{*}{$209$}                                \\
		M6                                                         & Response & {[}Networks, Routers, Ports, etc{]}                     &                                                     \\ \midrule
		\multicolumn{4}{c}{Nessus Scanning APIs}                                                                                                                                                   \\ \midrule
		M1                                                         & Request  & {[}User, Password{]}, {[}Nessus Session URL{]}          & \multirow{2}{*}{$471$}                                \\
		M2                                                         & Response & {[}Token{]}                                             &                                                     \\
		M3                                                         & Request  & {[}Token{]}, {[}Nessus Session URL{]}                   & \multirow{2}{*}{$4038$}                               \\
		M4                                                         & Response & {[}CVE-IDs, etc.{]}                                    & 									\\ \midrule
		\multicolumn{4}{c}{Diversity APIs: OS Diversification}  
		\\ \midrule
		M1                                                         & Request  & {[}Authentication Token{]}, {[}glance controller URL{]}          & \multirow{2}{*}{$559$}                                
		\\
		M2                                                         & Response & {[}List of images (OS instances){]}                                             &                                                     \\
		M3                                                         & Request  & {[}Authentication Token{]}, {[}VM ID, ImageRef, Nova{]}                   & \multirow{2}{*}{$18081^*$}                               \\
		M4                                                         & Response & {[}Status: accepted or abort{]}                                     &                                                     
		\\ \midrule
		\multicolumn{4}{c}{Redundancy APIs: VM Replicas}  
		\\ \midrule
		M1                                                         & Request  & {[}Token{]}, {[}Image\&FlavorRef, max\_count, Nova{]}                   & \multirow{2}{*}{$12091^*$}                               \\
		M2                                                         & Response & {[}Status: accepted or abort{]}                                     &                                                      
		\\ \midrule
		\multicolumn{4}{c}{Shuffle APIs: VM-LM}  
		\\ \midrule
		M1                                                         & Request  & {[}Auth. Token{]}, {[}VM ID, Host ID, Nova controller{]}                   & \multirow{2}{*}{$7216^*$}                               \\
		M2                                                         & Response & {[}Migration Status{]}                                     &                                                     \\ \bottomrule
	\end{tabular}
\end{table} 

\paragraph{Limitations} The update phase has not been implemented in cloud security framework. This includes running of Nessus scanning and recreation of HARM based on any changes captured in the cloud, such as updating VMs or vulnerabilities. We will further consider the update phase in our future work. Moreover, more in-depth evaluations for MTD techniques on real cloud infrastructure are needed such as performance analysis, measuring system downtime, \textit{etc}.

\section{Related Work}\label{sec:RW}
The theoretical investigation and evaluation of the security modeling and analysis adopting based on the MTD techniques for cloud computing have been proposed in the work~\cite{hong2016assessing, alavizadeh2017effective}. However, most of the proposed frameworks have focused on the implementations of GSMs on the networks~\cite{kotenko2013computer, dewri2012optimal, granadillo2012individual,jia2015towards}. The security modeling and analysis tools on the literature can be categorized based on the context of implementation test-bed such as cloud computing~\cite{chung2013nice}, networks and enterprises~\cite{kotenko2014evaluation}, or based on GSMs~\cite{hong2012harms}, ATs~\cite{dewri2012optimal}, AGs~\cite{ingols2009modeling, kotenko2013computer}, \textit{etc}., the automation approaches and levels~\cite{nespoli2018optimal}, or based on the effectiveness of solution like response time and the probability of success\cite{nespoli2018optimal}. The work~\cite{granadillo2012individual} proposed a prototype for 3D graphical visualization of the system, attack, and countermeasure model. The work~\cite{kotenko2014evaluation}proposed and implemented a fast network security assessment prototype based on the real scenario. However, the work~\cite{chung2013nice} developed a framework named NICE in the virtual network systems which is able to detect possible attacks against the cloud infrastructure. To the best of our knowledge, there is no prior work developing the MTD techniques incorporated with the automated GSMs in a cloud environment. In this paper, we developed an automated cloud security framework able to monitor and detect a private cloud and deploy MTD techniques on the infrastructures of the cloud.

\section{Conclusions}\label{sec:conclusion}
In this paper, we have investigated on practicability and usability of incorporating MTD techniques into GSMs as a framework on the real cloud.
We have developed a cloud security framework which is able to run on a private cloud platform named UniteCloud. The developed framework can 1) automatically monitor the cloud and collect the information such as hosts, VMs, network, and vulnerabilities existing on each VM using OpenStack APIs, 2) model and evaluate the cloud's security and adopt defensive MTD techniques, 3) automate the deployment of three MTD techniques OS Diversification as the Diversity technique, VM replication as the Redundancy technique, and VM-LM as the shuffle technique on the infrastructures of the UniteCloud using API calls, and 4) use a web application UI for interaction between the security experts and the backend engine of the framework and also visualize the generated security model.

\bibliographystyle{IEEEtran}
\bibliography{IEEEabrv,MTD}

\end{document}